# A Proactive Uncertainty driven Model for Tasks Offloading

**Maria Papathanasaki [a,b,1], Panagiotis Fountas [a,c], Kostas Kolomvatsos [a,d]**

[a] Department of Informatics and Telecommunications, University of Thessaly,

 Papasiopoulou 2-4, 35131, Lamia, Greece

[b] mpapathanasaki@uth.gr, [c] pfountas@uth.gr, [d] kostasks@uth.gr

**Abstract**

The ever-increasing demands of end-users on the Internet of Things (IoT), often cause great congestion in the nodes that serve their requests. Therefore, the problem of node overloading arises. In this article we attempt to solve the problem of heavy traffic in a node, by proposing a mechanism that keeps the node from overloading, regardless of the load entering in it, and which takes into consideration both the priority and the task demand. More specifically, we introduce a proactive, self-healing mechanism that utilizes fuzzy systems, in combination to a non-parametric statistic method. Through our approach, we manage to ensure the uninterrupted service of high demand or priority tasks, regardless of the load the node may receive, based on a proactive approach. Also, we ensure the fastest possible result delivery to the requestors, through the high priority and high demand sensitive mechanism. A series of experimental scenarios are used to evaluate the performance of the suggested model, while we present the relevant numerical results.

## 1. Introduction

A common challenge that the scientific community has to face today, is the overloading of nodes which leads to serving fewer high demand and high priority tasks. Node overloading happens when a server's resources are depleted to the point where it cannot handle incoming requests and thus it will not respond to them accordingly. In this article we aim to minimize the times that the queue fills up locally, in order to serve as many high priority and high demand tasks as feasible.

In the past, there have been various approaches to solve the network overloading problem. For instance, in [9] the authors study a M/M/1 queue subject to alternating behavior. They suppose that the queue's properties fluctuate randomly over time under two operational conditions, which allows to model queues based on two modes of customers arrivals, with fluctuating high-low rates. If the current environment leads to a traffic congestion, then the switch to the other environment may yield a favorable consequence for the queue length, achieving a stable queuing system constantly. In [21] a self-healing approach is proposed in order to maintain connectivity in reconfigurable networks.

---

[1] *Corresponding author*

In that paper, the case of the network being compromised by an attack is studied. The suggested distributed algorithm deletes the affected node and reconfigures the network structure in order to maintain its' functionality. This approach, however, is limited only to reconfigurable networks, in contrast to the approach we propose, which can be functional both in reconfigurable and non-reconfigurable networks. Both in [25] and in [6], a task offloading approach is introduced to manage the high load in mobile devices, by offloading computational and heavy tasks to nearby edge nodes for processing, in order to reduce the task processing delay to minimize the overall service time for latency sensitive (IoT) applications. In [19] the authors introduce a framework in which the tasks with high computing requirements are offloaded from the end devices to the edge nodes. An intelligent task offloading scheme that generates corresponding offloading decision profiles in varying scenarios is presented in [11]. This scheme combines machine learning and a learning-based offloading strategy based on historical data. In [26] the authors propose a low-complexity, greedy, heuristic algorithm to cope with the task offloading problem in a multi-cell Mobile-Edge Computing network.

In [27] the authors formulate task offloading using binary variables, in order to determine the best task offloading choices for mobile users. A model for simultaneous resource allocation and offloading decision optimization is suggested for mobile edge computing. A task offloading technique for mobile edge computing systems is described in [20] again. In contrast to prior research, the authors take a real-world scenario into consideration while optimizing the software runtime environment with task offloading. A reinforcement-learning based algorithm is proposed in [1], to deal with the resource management issue in the edge server and choose the best offloading strategy for reducing system costs, such as energy usage and computing time delay. In [2, 3, 5], the authors adopt the principles of Optimal Stopping Theory, to deal with the challenge of determining when to offload data to Edge Servers for computing analytics tasks. At the same time, they take into account the overall delay caused by each server. Furthermore, in [4], the authors extend the aforementioned works, considering the Quality of Service (QoS) assurance, and also, they reduce the anticipated execution time for analytics tasks. QoS is taken into consideration in [12] too. The authors introduce a heuristic algorithm to handle the offloading decision and assign priority to nodes that controls the order in which the nodes will perform the task offloading. Finally, in [22], a deep-reinforcement-learning algorithm is adopted. The purpose of the authors is to perform task offloading in Vehicular Fog Computing taking into account, in addition to other parameters, the task priority.

In contrast to the aforementioned approaches, the novelty of our paper is that the proposed model has the ability to keep enough space in the local queue, so as to serve the high priority tasks and tasks that exhibit a high demand, ensuring that the requestor will receive the answer to his query as soon as possible. At the same time, the low priority tasks are transferred in other nodes, in order to prevent the node from overloading. More specifically, our aim is to minimize the times that the queue fills up locally, in order to serve as many high priority and high demand tasks as possible.

The following list reports on the contributions of our work:

- We manage to serve as many high priority and high demand tasks as possible, by keeping always free space in the local queue especially for them.

- We ensure the smooth operation of the node under heavy traffic. More specifically, although the nodes that constitute the network have limited capacity, we have managed to ensure their seamless operation at most cases.

- We achieve the efficient management of load, by taking the appropriate decisions for the offloading actions, taking into consideration the task priority and demand.

- We report on the simulations used in the experimental assessment of the suggested model.

The rest of the paper is organized as follows. Section 2 outlines our problem and provides the main notations used in our model. Afterwards, in Section 3, we present the experimental results that were extracted out of the proposed mechanism. Finally, in Section 4 we conclude our paper by briefly summarizing the proposed mechanism and the benefits it offers. We also discuss the findings and their implications, as well as our future research plans.

## 2. Preliminaries and Scenario Description

### 2.1. Scenario Description

We consider a set of nodes $N = \{G_1, G_2, \ldots, G_n\}$ which interact among them and with their environment to collect data and execute various processing activities, i.e., tasks. The collected data that every $G_i$ locally stores are in the form of d-dimensional vectors i.e., $X^t = [x_1^t, x_2^t, \ldots, x_d^t]$ where the index $t$ depicts the time instance when $G_i$ receives $X_t$. This means that the recording of $X_t$ at the assumed discrete time instances $t \in \{1, 2, 3, \ldots\}$ defines a time series dataset upon which $G_i$ may be requested to execute various processing activities. Those requests can be initiated by users or applications indicating the execution of tasks upon the collected data and the extraction of knowledge. Apparently, for realizing the desired processing activities, $G_i$ adopts the local resources to perform the requested execution of tasks and store the outcomes. An additional step involves the communication with the requestor to deliver the final results. We assume that $G_i$ maintains a M/M/1/K queue, i.e., we consider a limited capacity of the local queue where the incoming tasks are stored before the local resources are allocated for their execution [24]. $G_i$ serves the incoming requests for processing in a First-Come-First-Served (FCFS) order to apply a 'fairness' in the allocation of resources. However, there are multiple scenarios where some tasks should have higher priority than others especially when we consider that $G_i$ should be able to support real time applications. For instance, imagine that $G_i$ is an edge node and should respond to a query that demands information about the traffic in the specific sub-area. $G_i$ should immediately respond to this query, otherwise, any delay may cause bottlenecks in the decision making relying on the outcomes of the processing with clear potential negative consequences in the execution of the supported applications.

$G_i$ does not want to be overloaded as it has to offload the incoming tasks to its peers causing additional delays in the provision of results. The research question taken into consideration in this paper is to assist $G_i$ in maintaining the size of the queue below a pre-defined threshold $\theta$ leaving room in the queue for hosting high priority tasks. Hence, a monitoring process should be applied that will result alerts when there is an increased risk to have $G_i$ overloaded, i.e., the number of tasks in the queue is above $\theta$. The discussed alerts will be, consequently, adopted to 'fire' a tasks management mechanism locally. Obviously, the scenario of having $G_i$ overloaded is met based on the rate of receiving the incoming tasks and the rate of executing them. In dynamic environments, these two rates are continually updated especially when we consider bursts of tasks reported in $G_i$. We propose the adoption of an uncertainty based reasoning mechanism to manage the uncertainty in the detection of $G_i$ as overloaded and the utilization of the aforementioned tasks management model to secure that there is always room for hosting high priority tasks in $G_i$ (the number of tasks in the queue should be less than $\theta$).

### 2.2. Tasks Management

The arrival of tasks in Gi is assumed to be governed by a Poisson process with rate λ. Based on this assumption, we can easily get the number of tasks Z(t) that arrive during the interval (0, t] that follows a Poisson distribution, described by the corresponding Probability Mass Function (PMF) [7]:

$$P[Z(t) = k] = \frac{e^{-\lambda t}(\lambda t)^k}{k!}, \quad k = 0, 1, 2, \ldots \quad (1)$$

Apparently, the inter-arrival time instances follows an exponential distribution with the respective Probability Density Function (PDF) given by the following equation [7].

$$f_I(x; \lambda) = \lambda e^{-\lambda x}, \quad x, \lambda > 0 \quad (2)$$

where λ represents the rate of arrivals. Under the same rationale, we consider that the time instances depicting the conclusion of the execution of tasks are governed by an Exponential distribution with the following PDF [7]:

$$f_S(x; \lambda) = \mu e^{-\mu x}, \quad x, \mu > 0 \tag{3}$$

where μ depicts the service rate of $G_i$. In the above discussed PDFs, we adopt an index (I or S) to depict the distribution that corresponds to the arrival or the execution of tasks, respectively. Based on the previous assumptions, we can easily observe that the average inter-arrival time and the average service time is given by the following equations [7]:

$$\Omega[I] = \frac{1}{\lambda} \tag{4}$$

$$\Omega[S] = \frac{1}{\mu} \tag{5}$$

Equation (6) represents the traffic intensity in a M/M/1/K system given by the ratio of the arrival rate, compared to the service rate. The following equation stands true [7, 24]:

$$\rho = \frac{\lambda}{\mu} \tag{6}$$

With simple calculations, we can extract the probability of having v tasks in Gi's local queue, as the following equation exposes [7]:

$$P_v = \begin{cases} \frac{(1-\rho)\rho^v}{1-\rho^{K+1}}, & \rho \neq 1, \quad 1 \leq v \leq K \\ \frac{1}{K+1}, & \rho \neq 1, \quad 1 \leq v \leq K \end{cases} \tag{7}$$

If we assume that θ tasks should be at most in the local queue, we can easily calculate the probability of overloading $P_\delta$ as follows [7]:

$$P_\delta = \sum_{v=\theta}^{K} P_v \tag{8}$$

θ is the threshold for considering $G_i$ as overloaded and start offloading the upcoming tasks to peer nodes. For instance, offloading can be performed through the adoption of various techniques like [17, 23, 14, 15, 16]. With the proposed model, we are able to support $G_i$ to maintain a free room for high priority tasks, based on a proactive approach. We have to notice that the processes upon which the offloading decisions and the selection of the appropriate peers to host the offloaded tasks take place, lies beyond the scope of this paper [10].

## 2.3. Trend Estimation

Let us consider discrete time intervals where we record the inter-arrival rate of tasks and their execution rate. Obviously, if $\lambda \gg \mu$, $G_i$ is not stable, it will certainly be overloaded. However, in real setups, we can observe fluctuations in the realization of $\lambda$ and $\mu$. Hence, we decide to focus on every discrete time interval (e.g., minutes, hours, days, etc) and deal with the rate of execution μ as a time series dataset. Our target is to check the trend of the execution rate, in order to support the decision making concerning the offloading of tasks. In simple words, if we see that $\mu$ increases, we can postpone the offloading of tasks to eliminate the time required for sending tasks to peers and getting the final results to deliver them to the requestor. For instance, new tasks placed for execution may request simple processing activities or actions that can be based on previous results, thus, if we re-use them, we can significantly reduce the execution time. We rely on the time series $\{\mu_1, \mu_2, \ldots\}$ where $\mu_j$ is the execution rate at each discrete interval and apply a simple, however, powerful technique for extracting the trend of the execution rate. This technique is the Kendall's tau statistic [18, 8].

Kendall's tau ($\tau$) coefficient is a non-parametric statistic method, which is used to measure the correlation and the relationship strength between two measured quantities. This coefficient is applied upon the aforementioned $\mu$ time series $\{\mu_1, \mu_2, \ldots\}$. $\tau$ value is bounded in $[-1, 1]$ and calculated by the following equation:

$$\tau = \frac{S}{\binom{n}{2}} \quad (9)$$

where n is the number of μ upon which the Kendall's tau is applied and S depicts the Kendall Score which is computed as follows:

$$S = \sum_{i=1}^{n-1} \sum_{j=i+1}^{n} sign(\mu_j - \mu_i) \quad (10)$$

$$sign(\mu_j - \mu_i) = \begin{cases} 1, & \mu_j - \mu_i > 0 \\ 0 & \mu_j - \mu_i = 0 \\ -1 & \mu_j - \mu_i < 0 \end{cases} \quad (11)$$

The closer to $+1$ the value of $\tau$, the more increasing the trend is, while the most decreasing trend is achieved when $\tau$ is closer to $-1$. The data have no trend through the time, when $\tau$ is not significantly different from zero. In our model, we normalize the value of $\tau$ in range $[0, 1]$.

### 2.4. Uncertainty-based Estimation of Overloading

In our model, we are based on two parameters in the examined time interval to take a decision if a node tends to become overloaded or not. The first parameter is the value of $\tau$, which shows the trend of μ through the time intervals. The second parameter is the probability $P_\delta$, which depicts that the fullness of memory exceeds a threshold $\theta$. The realizations of $P_\delta$ and $\tau$, are used as inputs to define the antecedent part in our Mamdani fuzzy system to set a value for the consequent part i.e., the Overload Indicator ($OI$) which is bounded in $[0, 1]$. We define the fuzzy knowledge base, every time a node receives a request for the service of a task, e.g., a set of Fuzzy Rules (FRs) in the form of: 'When the probability the fullness of the memory exceeds $\theta$ is high and the trend of service rate is high negative, the $OI$ for that node will be extreme high'. The proposed FRs have the following structure:

**IF $P_\delta$ is $A_1$ AND $\tau$ is $A_2$ THEN $OI$ is $B$,**

where $A_1$, $A_2$ and $B$ are the membership functions for the FRs mapping. For Fuzzy Logic (FL) set, we characterize the value of $A_1$ and $B$ by the terms extreme low, low, medium, high and extreme high and the value of $A_2$ through the terms high negative, low negative, neutral, low positive, high positive. In our Mamdani fuzzy system, we adopt triangular membership functions for the definition of bounds for every term of $P_\delta$, $\tau$ and $OI$ and we involve the centroid defuzzification method for the calculation of the final crisp output value i.e., the value of $OI$. We have to notice that

FRs and membership functions for the proposed Mamdani fuzzy system, are defined by the experts. We consider a knowledge base with 25 rules, i.e., a rule for every combination of the two inputs.

*2.5. Offloading Strategy based on the Overloading Indicator*

Assume that the outcome of the Fuzzy System, i.e., the $OI$ indicates that the local queue will be overloaded and offloading actions should take place. We can consider the following cases. **Case A**: the number of tasks in the local queue is grater or equal to $\theta$ (our mechanism delayed to detect the overloading status); **Case B**: the number of tasks in the local queue is less than $\theta$ (our mechanism proactively detects the overloading status). In Case A, we select to offload all the upcoming tasks to peers and high priority tasks will be locally placed if there is room for that. In Case B, we propose the use of an additional mechanism that selects to keep some tasks locally till the threshold $\theta$ as follows. High priority tasks and tasks that exhibit a high demand locally, are kept in the queue only if we do not violate $\theta$. A study on how the demand for tasks may affect their management can be found in [13]. Hence, we can benefit from reusing previously calculated results, spending less resources and maximizing the node performance. The proposed processing is depicted by the Algorithm 1.

---

**Algorithm 1** Algorithm for Tasks Offloading based on $OI$

---

**while** True **do**
    Receive a task
    $OI = FuzzySystem()$
    **if** Queue is full **then**
      Offload the task
    **else**
      **if** $OI > \beta$ **then**                                                     ▷ $\beta$ is a pre-defined threshold
        **if** Number of local Tasks $\geq \theta$ **then**                                                 ▷ Case A
            **if** Task has high priority **then**
                Place the task in the local queue
            **else**
                Offload the task
            **end if**
        **else**                                                                                            ▷ Case B
            **if** Task has high priority OR high demand **then**
                Place the task in the local queue
            **else**
                Offload the task
            **end if**
        **end if**
      **end if**
    **end if**
**end while**

---

## 3. Results

*3.1. Setup and Performance Metrics*

The experimental evaluation of the Uncertainty driven Proactive Self Healing Model (UPSHM), relies on the Tasks Simulation Dataset[2] and the comparison to a Baseline Model (BM). The aforementioned dataset contains $E = 3564$ instances, which represent tasks that arrive in a node in $W = 100$ time intervals. More specifically, every instance has five attributes; the id of the task, the service time of the task, the arrival time, the priority value and the demand value. The number of tasks in each time interval, follows a Poisson distribution and the service and arrivals times of the tasks, follow an exponential distribution. In addition, the priority and the demand values have an uniform distribution. For every time interval, we generate randomly the values of $\lambda$ and $\mu$, based on uniform distribution in range $[10, 60]$ and $[10, 40]$ respectively. For comparison purposes, we define a BM which performs an offloading action when the local memory of a node is full, regardless the priority or the demand of a task. In the comparison between the performance of UPSHM and BM, we pay attention on the number of tasks that the node executed locally. This quantity is expressed by the $\phi$ metric, which is defined by the Equation (12).

$$\phi = \frac{\xi}{E} \tag{12}$$

where $\xi$ is the number of executed tasks and $E$ is the total number of tasks that arrive in the node, during a number of time intervals. The higher the number of the tasks that are executed locally, the lower the delays that causing in the provision of results is. Additionally, we examine the performance of both BM and UPSHM on the percentage of high priority tasks that were executed locally. The following equation defines the aforementioned metric

$$\eta = \frac{\kappa}{L} \tag{13}$$

where $\kappa$ is the number of high priority tasks which were executed locally and $L$ is the total number of high priority tasks that arrive in the node, during a number of time intervals. When $\eta \to 1$ it is indicated that the node executed all the high priority tasks locally, delivering the results to the requestor in the minimum possible time. The same scenario stands for the tasks with high demand rates, where $\omega$ is the number of tasks that were executed locally. The next equation holds true:

$$\omega = \frac{\zeta}{\gamma} \tag{14}$$

where $\zeta$ is the number of high demand tasks which were executed locally and $\gamma$ is the total number of tasks with high demand. The ability of both UPSHM and BM to maintain free space for the high priority tasks, is described by the $\psi$ as follows:

$$\psi = \frac{q}{E} \tag{15}$$

where $q$ is the total number of times that the local memory was full. When $\psi \to 1$, it is indicated that the node offloads tasks uncontrollably, without taking into consideration the priority or the demand of the tasks. We perform simulations for the following values of thresholds; $\beta = \{0.8, 0.84, 0.88\}$ and $\theta = \{85\%, 90\%, 95\%\}$

3.2. *Performance Assessment*

---

[2] http://www.iprism.eu/assets/DatasetTasksOffloading.zip

In Figure 1, we present the results of the experiment for the metrics φ and η for the values of thresholds $\theta = 85\%$ and $\beta = 0.8$. In these results, we notice that the BM executed higher percentage of tasks, in contrast to UPSHM. However, when we focus on the percentage of high priority tasks that were executed locally, we can easily come to the conclusion that the proposed model clearly outperforms the BM by executing all high priority tasks locally. Additionally, the Figure 2 shows the models' performance for metrics $\omega$ and $\psi$. As far as the percentage of tasks with high demands is concerned, the UPSHM has better performance executing almost all of these tasks. In the last metric i.e., $\psi$ the ability of our model to maintain free space in the local memory for high priority tasks is indicated, since the memory of the node had not been full during the W time intervals. Summarizing, from the results of the first experiment we can clearly observe the significant dominance of our model in almost all the metrics.

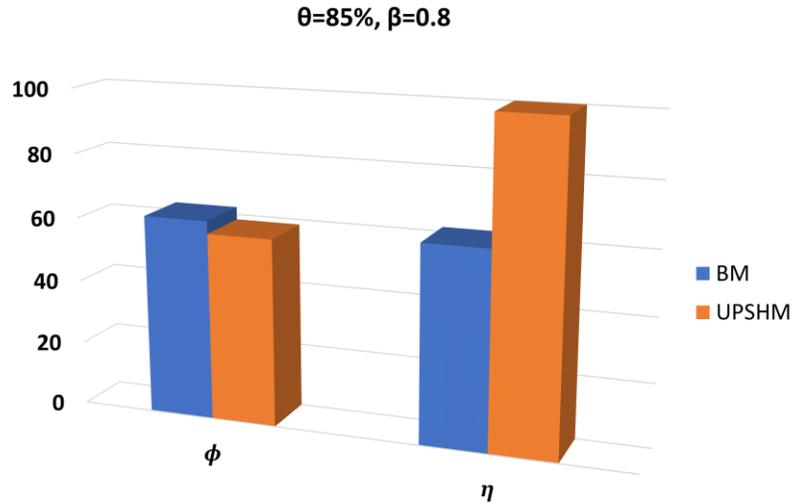

Figure 1: Comparison of results for $\phi$ and $\eta$ metric.

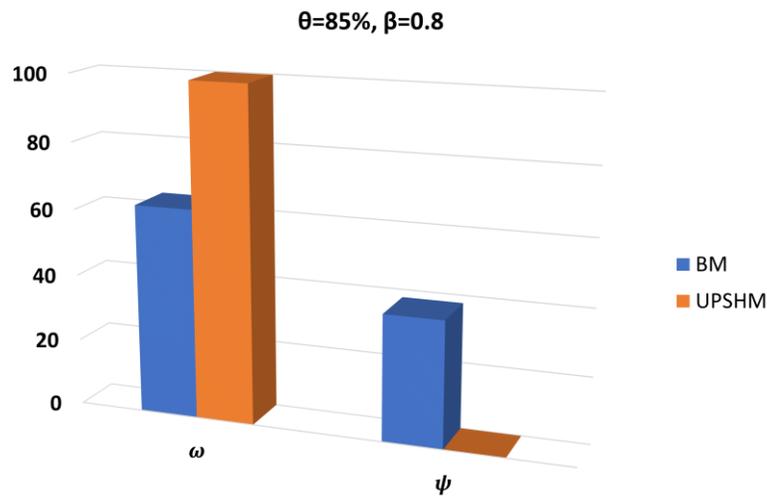

Figure 2: Comparison of results for ω and $\psi$ metric.

In the second experimental scenario we set the $\beta$ threshold at 0.88 and we keep $\theta$ static at 85%. Figure 3, depicts the performances of both BM and UPSHM for metrics φ and η. We notice that the BM executed slightly lower percentage of received tasks than the UPSHM. Nevertheless, when we examine the percentage of high priority tasks that were executed locally, we can easily perceive that the difference in the performances between the BM and UPSHM, is noticeably high. In Figure 4 we see the behavior of the models in the execution of tasks with high rate of demand and also the number of times that the local memory ran out of space for all types of tasks. We can easily come to the conclusion that the UPSHM has better performance for the $\omega$ metric, by executing more tasks with high demand in contrast to the BM. Additionally, in $\psi$ metric, we notice that both UPSHM and BM exhausted the capacity of the local node memory several times. This means that in this specific experimental scenario, regardless the model that we used for the management of the incoming tasks, the node was forced to offload the incoming tasks without considering their priority or their demand from others peers. Even in this case though, the UPSHM outperforms the BM in $\psi$ metric, because the number of the times that the local node memory was full, was significantly less than BM. In conclusion, the results of this experimental set, depict the advantage of the UPSHM in all metrics.

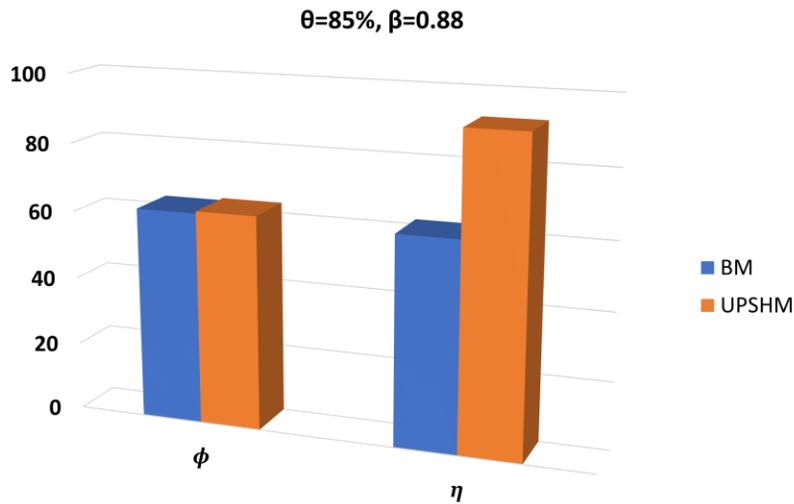

**Figure 3: Comparison of results for φ and η metrics.**

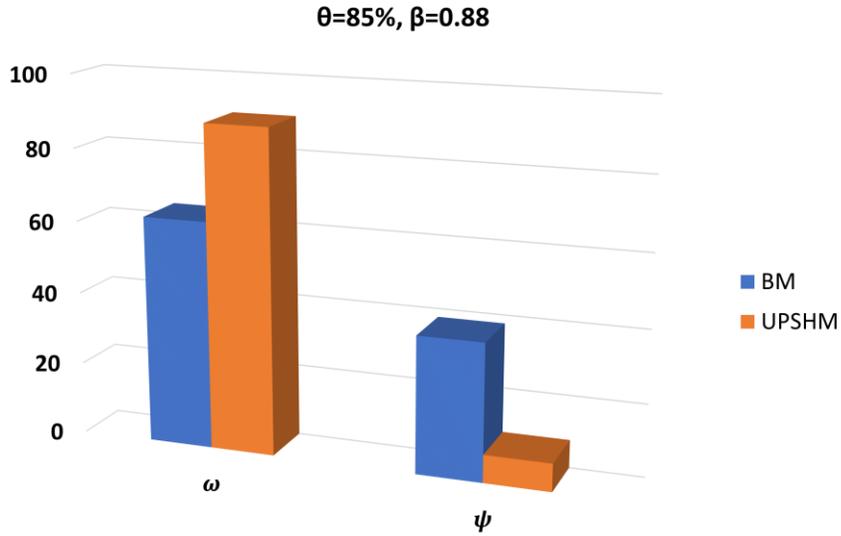

Figure 4: Comparison of results for φ and η metric.

In Figure 5 we present a comparison of the performances of UPSHM for different values of β threshold and $\theta = 85\%$. We can plainly see by the values of φ metric, the higher the value of $\beta$ threshold, the more tasks are executed locally in a node. As the number of tasks with high priority or high demand is concerned, the values of metrics $\eta$ and ω depict that they are affected in opposite direction to $\phi$, as we notice a slight decrease in their values. When we focus on the number of times that the node memory was out of space, we perceive that it is increased when we increase the $\beta$ threshold, as the $\psi$ metric shows. In Figure 6 a comparison of the performances of UPSHM for different values of $\theta$ threshold and $\beta = 0.8$ is presented. Same as previous the values of $\phi$ and $\psi$ metrics increase as $\theta$ increases. On the contrary, the values of metrics η and $\omega$ are affected in an adverse direction.

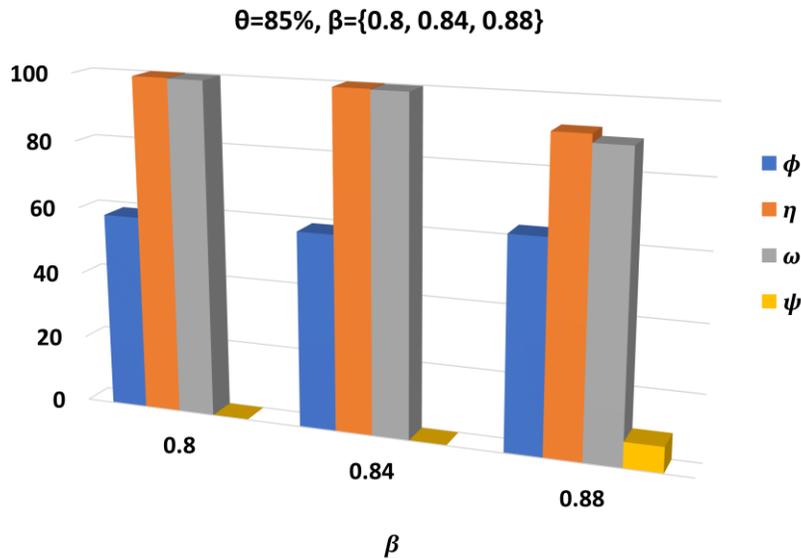

Figure 5: Comparison of UPSHM performances for all metrics for different values of β threshold.

Observing the Figure 5 and Figure 6, we can extract conclusion for which of the metrics $\theta$ and $\beta$ affects more the performance of UPSHM. We see that $\phi$ metric has almost the same performance in both figures. When we focus on metrics η, ω and ψ we notice that both thresholds affect the performance of UPSHM negatively. However an increase in $\beta$ threshold has greater negative effect than the increase in $\theta$ threshold.

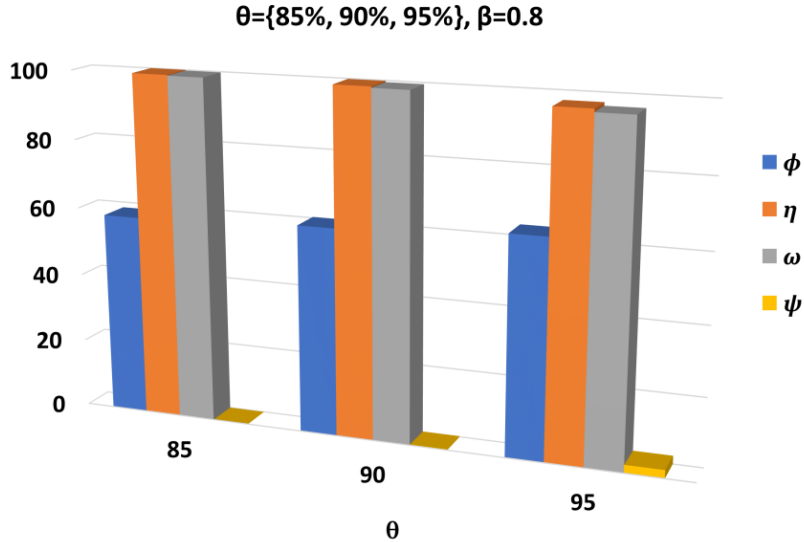

Figure 6: Comparison of UPSHM performances for all metrics for different values of θ threshold.

## 4. Conclusions

A significant research subject is the preventing of the node overloading, which leads to uncontrollable task offloading to peer nodes. Some of these offloaded tasks, might be in high demand or high priority and thus, efficient mechanisms that allow serving more of these tasks, should be created. In this paper, we approach this problem by developing a mechanism that makes a node capable of handling a large volume of tasks without being critically overloaded for large quanta of time. This way, we achieve to serve high percentage of tasks with high priority and demand, since we always try to keep space for them locally in the node's queue, based on a proactive approach. In the proposed model, we adopt an uncertainty based reasoning mechanism, to manage the uncertainty in the detection of the node as overloaded and the utilization of a monitoring process, to secure that there is always enough free space for hosting high priority and high demand tasks in the node. We evaluate the performance of our model through various experiments, and we compare it with a baseline method. Our experimental evaluation, demonstrates that the suggested approach can effectively help to achieve the desired outcomes, which are supported by numerical data. Our future research directions involve, the definition and adoption of a more complex methodology for making a decision about when a node is considered as overloaded and when an offload action should take place. Furthermore, an additional future direction is the study of the mobility of the users, in combination to the tasks that they render as high demand.